\documentclass[conference,10pt,bottom=0.95in]{IEEEtran}
\IEEEoverridecommandlockouts
\IEEEoverridecommandlockouts
\usepackage{cite,comment}
\usepackage{amsmath,amssymb,amsfonts}
\usepackage{algorithmic}
\usepackage{graphicx}
\usepackage{textcomp}
\usepackage{xcolor}
\usepackage{soul}
\def\BibTeX{{\rm B\kern-.05em{\sc i\kern-.025em b}\kern-.08em
    T\kern-.1667em\lower.7ex\hbox{E}\kern-.125emX}}
    \usepackage[letterpaper,top=0.7in,bottom=0.95in,left=0.625in,right=0.625in]{geometry}

\begin{document}

\title{Neurosymbolic Learning for Advanced Persistent Threat Detection under Extreme Class Imbalance\\
\vspace{-6mm}
\thanks{This work is supported by Commonwealth Cyber Initiative under Grant No. 301108-010 }
}
\vspace{-5mm}\author{
\IEEEauthorblockN{
Quhura Fathima\IEEEauthorrefmark{1},
Neda Moghim\IEEEauthorrefmark{2}\IEEEauthorrefmark{1},
Mostafa Taghizade Firouzjaee\IEEEauthorrefmark{3},
Christo K. Thomas\IEEEauthorrefmark{4},
Ross Gore\IEEEauthorrefmark{1},
Walid Saad\IEEEauthorrefmark{5}
}
\IEEEauthorblockA{\IEEEauthorrefmark{1}Center for Secure \& Intelligent Critical Systems and School of Cybersecurity, Old Dominion University, VA, USA}

\IEEEauthorblockA{\IEEEauthorrefmark{2}Department of Computer Engineering, University of Isfahan, Iran}

\IEEEauthorblockA{\IEEEauthorrefmark{3}Faculty of Engineering Modern Technologies, Amol University of Special Modern Technologies, Iran}

\IEEEauthorblockA{\IEEEauthorrefmark{4}Department of Electrical and Computer Engineering, Worcester Polytechnic Institute, MA, USA}

\IEEEauthorblockA{\IEEEauthorrefmark{5}Department of Electrical and Computer Engineering, Virginia Tech, VA, USA}

\IEEEauthorblockA{
Emails: \{qfath001, nmoghim, rgore\}@odu.edu, 
n.moghim@eng.ui.ac.ir, 
taghizade.mostafa@gmail.com,\\
cthomas2@wpi.edu, 
walids@vt.edu
}
\vspace{-0.4cm}}

\maketitle
\begin{abstract}
The growing deployment of Internet of Things (IoT) devices in smart cities and industrial environments increases vulnerability to stealthy, multi-stage advanced persistent threats (APTs) that exploit wireless communication. Detection is challenging due to severe class imbalance in network traffic, which limits the effectiveness of traditional deep learning approaches and their lack of explainability in classification decisions. To address these challenges, this paper proposes a neurosymbolic architecture that integrates an optimized BERT model with logic tensor networks (LTN) for explainable APT detection in wireless IoT networks. The proposed method addresses the challenges of mobile IoT environments through efficient feature encoding that transforms network flow data into BERT-compatible sequences while preserving temporal dependencies critical for APT stage identification. Severe class imbalance is mitigated using focal loss, hierarchical classification that separates normal traffic detection from attack categorization, and adaptive sampling strategies. Evaluation on the SCVIC-APT2021 dataset demonstrates an operationally viable binary classification F1 score of $95.27\%$ with a false positive rate of $0.14\%$, and a $76.75\%$ macro F1 score for multi-class attack categorization. Furthermore, a novel explainability analysis statistically validates the importance of distinct network features. These results demonstrate that neurosymbolic learning enables high-performance, interpretable, and operationally viable APT detection for IoT network monitoring architectures.
\end{abstract}
\vspace{-2mm}\section{Introduction}\vspace{-1mm}
Widespread Internet of Things (IoT) deployment across smart cities and critical infrastructure revolutionizes connectivity but simultaneously expands attack surfaces for 
sophisticated adversaries \cite{Ogbodo2022}. Modern IoT deployments across diverse wireless protocols (LPWAN to 5G) present attack vectors that traditional IDS struggle to address \cite{Proto2025}. Unlike centralized enterprise networks \cite{Lyu2024}, wireless IoT environments are distributed across heterogeneous devices, requiring security approaches adapted to diverse network architectures \cite{Canavese2024}. Advanced persistent threats (APTs) in wireless IoT represent an insidious threat class \cite{Popoola2025}: adversaries exploit wireless patterns and device vulnerabilities to establish long-term presence through coordinated multi-stage campaigns including initial compromise, reconnaissance, lateral movement, pivoting, and data exfiltration. This distributed architecture makes detection difficult: attackers blend malicious data with legitimate traffic while gradually expanding their foothold. APT detection in wireless IoT networks faces two interrelated challenges. First, \emph{extreme class imbalance} (e.g., $98\%$ benign traffic) makes attack samples extraordinarily rare, limiting conventional machine learning (ML) approaches that optimize for overall accuracy. Second, existing systems remain \emph{opaque ``black boxes,''} preventing analysts from understanding attack patterns, validating alerts, or developing mitigation strategies, which are capabilities essential for autonomous deployment with unavailable human oversight.
\vspace{-3mm}\subsection{Related Works}\vspace{-2mm}
ML-based intrusion detection system (IDS) approaches \cite{Sun2020,sinha2025high,Chawla2002,Hussain2024,YangPeng2025} optimize for accuracy or efficiency at the cost of transparency. Hybrid CNN-LSTM architectures \cite{Sun2020,sinha2025high} achieve $99.87\%$ accuracy on BoT-IoT, while ensemble approaches \cite{Chawla2002,Hussain2024} combine SMOTE-based balancing with tree ensembles, reporting $99.74\%$ accuracy on UNSW-NB15. However, these techniques generate synthetic samples that obscure genuine attack patterns and provide no grounded explanations. 
 Transformer-based IDS approaches, including IDS-INT \cite{Ullah2023}, which combines transformer-based transfer learning with SMOTE and CNN, and BERT-based EBIDS \cite{Sattarpour2025}, achieve high accuracy on Edge-IIoTset and CICDos 2017 through multi-layer detection, outperforming CNN/LSTM methods. Contemporary hybrid ML and large language model (LLM) frameworks integrate separate detection and reasoning components for post-hoc behavioral analysis and mitigation suggestions. While transformer-based models \cite{Devlin2018, hu2024edge} show promise for pattern recognition, they assume sufficient training data and typically require deployment on servers with adequate computational resources.  All these approaches \cite{Sun2020,sinha2025high,Chawla2002, Hussain2024,YangPeng2025,Ullah2023,Sattarpour2025, Devlin2018, hu2024edge} remain ``black boxes'' providing no interpretability of decisions, 
 limiting applicability to autonomous systems requiring transparent reasoning.   Recent work \cite{khediri2024shap} and \cite{mutalib2024explainable} on explainable ML for IDS  has explored shapley additive Explanations (SHAP) and local interpretable model-agnostic explanations (LIME) for post-hoc explanation of IDS models. While effective for transparency, these post-hoc approaches face a fundamental limitation: explanations are computed after training and may not reflect the actual decision pathways.  

\vspace{-2mm}\subsection{Contributions}\vspace{-1.5mm}
The main contribution of this paper is a neurosymbolic framework unifying BERT and LTNs to achieve explainable APT detection in class-imbalanced IoT networks with quantified statistical validation, which is the first in transformer-based IoT IDS literature. Contrary to post-hoc explanation methods \cite{khediri2024shap} and \cite{mutalib2024explainable}, this work advances the field by integrating explainability into training through BERT attention weights and LTN logical predicates, ensuring explanations are grounded in genuine attack signatures rather than learning artifacts. Our key contributions are:
\begin{itemize}
\item We develop a neurosymbolic framework that integrates BERT's transformer-based pattern recognition (neural component) with LTN's logical reasoning constraints (symbolic component) during training. This produces detection results grounded in explicit domain intepretable concept, and hence are explainable.
\item We develop a hierarchical two-stage classification that separates binary detection (benign vs attack) from multi-class APT categorization, addressing extreme class imbalance at the architectural level. This approach avoids synthetic sample ambiguity inherent in SMOTE pipelines while preserving the interpretability of learned patterns.
    \item We provide rigorous statistical evidence that $75\%$ of extracted features show significant differences between attack and normal traffic. This quantitative validation, which, to our best knowledge, is the first in transformer-based IoT IDS, demonstrates that explanations reflect genuine attack signatures, enabling transparent autonomous deployment. 
\end{itemize}
\vspace{-3mm}\section{Model and Methodology} \vspace{-1mm}Our goal is to detect and classify multi-stage APTs from network flow data in an IoT environment characterized by extreme class imbalance.
\vspace{-2mm}\subsection{Dataset and Preprocessing} \vspace{-1mm} We use the SCVIC-APT2021 dataset \cite{liu2022new}, which provides realistic IoT network traffic and exhibits the severe class imbalance (98.35\% benign) central to our problem, as shown in Table \ref{tab:class_distribution}. While benchmark datasets like KDD Cup 99 \cite{PervezSKIMA2014}, NSL-KDD, and CICIDS \cite{YuliantoJP2019} are commonly used, they often have shortcomings for IoT security evaluation, such as outdated attacks and synthetic traffic distributions not representative of modern IoT deployments. The SCVIC-APT2021 dataset, by contrast, was developed with a focus on realistic, modern network architectures and APT patterns relevant to IoT environments. It includes over $315,000$ records and was initially composed of $84$ features capturing rich network telemetry suitable for IoT traffic analysis. Each network flow sample $\boldsymbol{x}$ represents a vector of features capturing flow-level statistics (flow duration, packet sizes), behavioral patterns (inter-arrival times, TCP flags), and endpoint characteristics (ports) extracted from network communications in IoT-enabled environments.
\begin{table}[t]
\vspace{-0mm}\centering
\caption{Class distribution in SCVIC-APT2021 dataset }\vspace{-2mm}
\label{tab:class_distribution}
\begin{tabular}{lrr}
\hline
\textbf{Class} & \textbf{Samples} & \textbf{Percentage} \\
\hline
Normal Traffic & 254,836 & 98.35\% \\
Pivoting & 2,122 & 0.82\% \\
Reconnaissance & 833 & 0.32\% \\
Lateral Movement & 729 & 0.28\% \\
Data Exfiltration & 527 & 0.20\% \\
Initial Compromise & 73 & 0.03\% \\
\hline
\end{tabular}
\vspace{-6mm}\end{table}
Standard preprocessing removed duplicates and corrected synchronization errors, while data quality analysis revealed integrity issues requiring specialized handling for IoT traffic characteristics. Missing 'Flow Bytes/s' values were set to zero for flows with zero duration (common in IoT signaling traffic). Four corrupted timing attributes were removed, reducing features from $84$ to $80$. We applied feature-aware normalization based on IoT traffic distributional characteristics. After discarding non-predictive columns, the remaining $75$ features were categorized as: (a) $18$ binary flags and categorical identifiers maintained original scales; (b) $9$ features with outliers (e.g., flow rates) normalized via median centering and interquartile range scaling; and (c) $48$ remaining features (e.g., ports) standardized to zero mean and unit variance. Consensus-based feature selection combined four diverse statistical approaches: random forest (non-linear interactions), extra trees (ensemble validation), mutual information (non-parametric dependencies), and F-test (linear discriminative power). These four methods represent fundamentally different evaluation criteria, ensuring features rank high across multiple methods rather than appearing important to only one. We retained features consistently ranked high across all four. Discriminative power was validated via t-test ($p<0.001$, Cohen's $d>0.3$) \cite{cohen2013statistical}. Analysis of data exfiltration misclassification patterns identified two additional features: backward IAT total and total backward packets. The final $12$-feature set balances discriminative power with efficient feature representation. These $12$ features feed both neural and symbolic architecture components, which are discussed next.

\vspace{-2.5mm}\subsection{BERT-LTN Neurosymbolic Architecture} \vspace{-1mm} Our architecture, shown in Fig.1, integrates a BERT-based neural path for 
pattern recognition and an LTN-based symbolic path for logical reasoning. Both 
paths process the same $12$ network flow features extracted during preprocessing.
\subsubsection{Neural component using BERT}
For the neural path, we adapt a pre-trained bert-base-uncased model \cite{Devlin2018}. BERT expects sequential token inputs, but network traffic is inherently tabular. 
We bridge this gap by encoding each of the 12 features as a learned dense representation. 
Specifically, each feature $x_i$ is projected independently into BERT's 768-dimensional 
hidden space via:
$
\boldsymbol{e}_i = \boldsymbol{w}_i x_i + \boldsymbol{b}_i,
$
where $\boldsymbol{w}_i \in \mathbb{R}^{768}$ and $\boldsymbol{b}_i \in \mathbb{R}^{768}$ are learnable parameters. 
This produces $12$ feature tokens. We append two special tokens, [CLS] at the beginning 
(used for sequence-level classification) and [SEP] at the end (marking sequence boundaries), creating 
a $14$-token sequence. This design preserves the semantic meaning of each network feature 
(e.g., ``packet size'' remains interpretable as a specific dimension) while enabling BERT's 
multi-head attention mechanisms to \emph{learn complex interactions among flow statistics, 
temporal patterns, and protocol behaviors} without assuming any predefined relationships 
among features. The multi-head attention mechanism in BERT's $k$ transformer layers produces attention 
heads $h \in [1, h_{\max}]$ that learn different interaction patterns. 
For each flow sample $\boldsymbol{x}$, we extract per-token attention weights 
$\boldsymbol{\alpha} \in \mathbb{R}^{14}$ from different layers and heads, measuring 
how much each feature token contributes to the final [CLS] representation. These 
attention weights provide the first explainability mechanism called \textit{feature attribution}. 
Specifically, for the [CLS] token at layer $l$ and head $h$: $
\boldsymbol{\alpha}^{(l,h)} = 
\text{softmax}\big([a_1^{(l,h)}, a_2^{(l,h)}, \ldots, a_{14}^{(l,h)}]\big),
$
where $a_i^{(l,h)}$ denotes the attention score of token $i$ in layer $l$, head $h$. 
The attention over feature tokens (tokens $1–12$) provides a ranking of feature 
importance for each prediction. By aggregating attention weights across layers and 
heads, we obtain per-feature importance scores as
$
\propto \sum\limits_{l,h} \alpha_i^{(l,h)}.
$
This formulation grounds explainability directly in BERT’s learned representations, 
as the model explicitly learns which features are most influential for each prediction.

\begin{figure}[t]
\centering
\includegraphics[width=1\columnwidth]{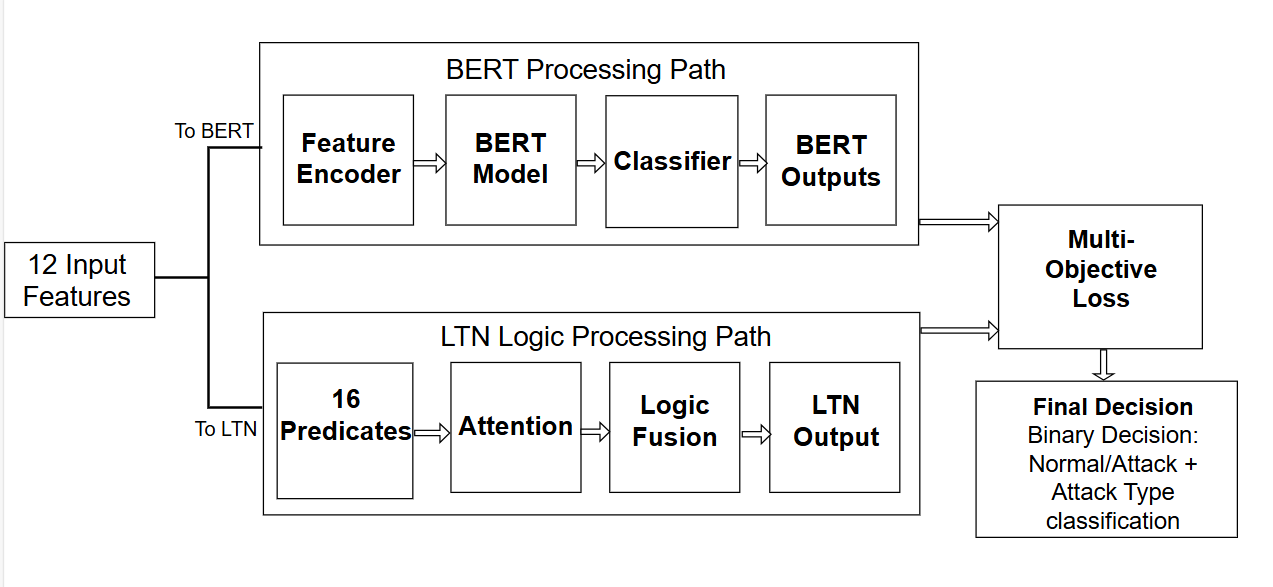}
\vspace{-7mm}\caption{System architecture showing parallel BERT and LTN pipelines, hierarchical classification heads, and multi-objective training integration.}
\label{fig:architecture}
\vspace{-6mm}\end{figure}

\subsubsection{Symbolic component using LTN}In parallel, the symbolic path operates directly on the $12$ normalized features 
to provide interpretable logical reasoning. The LTN component implements $16$ 
learnable logical predicates, each realized as a $3$-layer MLP with architecture 
$\mathbb{R}^{12} \rightarrow \mathbb{R}^{32} \rightarrow \mathbb{R}^{16} \rightarrow \mathbb{R}^{1}$: 
two hidden layers with ReLU activation (dimensions $12 \rightarrow 32 \rightarrow 16$) 
followed by a sigmoid output layer. Each predicate captures a \emph{domain-interpretable 
concept} (e.g., ``large forward data transfer'', ``unusual port activity'', ``high packet rate'') 
and outputs a satisfaction degree in the range $[0, 1]$. For each predicate $i$, we compute an attention-weighted input $\tilde{x}_i$ by applying 
learnable attention weights $\boldsymbol{a}_i = [a_{i1}, a_{i2}, \ldots, a_{i12}] \in \mathbb{R}^{12}$ 
to the 12 normalized input features $\boldsymbol{x} = [x_1, x_2, \ldots, x_{12}]$,
\vspace{-6pt}
\begin{equation}\tilde{\boldsymbol{x}}_i = \boldsymbol{a}_i \odot \boldsymbol{x} = [a_{i1}x_1,\, a_{i2}x_2,\, \ldots,\, a_{i12}x_{12}], \tag{1a}\end{equation}
where $\odot$ denotes element-wise multiplication. For each predicate $i \in \{1, \cdots, 16\}$, we learn predicate-specific attention weights 
that determine which input features are most relevant. We parameterize this via a 
learnable attention vector $\boldsymbol{v}_i \in \mathbb{R}^{12}$, which is optimized during training via 
backpropagation. These attention weights determine 
which features are most relevant for each predicate. For each predicate $i$, the attention vector is computed as:
\vspace{-2mm}\begin{equation}
   \mathbf{a}_i = \text{softmax}(\mathbf{v}_i), \quad 
\text{where} \quad 
a_{i,j} = \frac{\exp(v_{i,j})}{\sum_{k=1}^{12} \exp(v_{i,k})}.  
\vspace{-1mm}\end{equation}
This produces normalized attention weights $\mathbf{a}_i \in [0,1]^{12}$ satisfying 
$\sum_j a_{i,j} = 1$. These weights measure the importance of each feature for 
predicate $\pi_i$: a high $a_{i,j}$ indicates that feature $x_j$ is important for 
predicate $\pi_i$, while a low $a_{i,j}$ indicates suppression. The raw attention 
parameters $\mathbf{v}_i$ are learned during training, where features with large $v_{i,j}$ 
values receive higher attention after the softmax operation, whereas those with 
smaller $v_{i,j}$ values receive lower attention.
 For example, the predicate 
``large forward data transfer'' learns high attention weights for features such as 
``total length of Fwd Packet'' and ``Fwd packet length max'', while suppressing attention 
to irrelevant features like port numbers.
Further, the satisfaction degrees for all predicates 
are combined using learnable importance weights $w_i$ as follows:
\vspace{-2mm}\begin{equation}
   \phi = \sigma\left(\sum_{i=1}^{16} w_i \pi_i(\tilde{\boldsymbol{x}}_i)\right), \label{eq_satis_deg}\vspace{-1mm} 
\vspace{-1mm}\end{equation}
where $\pi_i(\tilde{\boldsymbol{x}}_i) \in [0, 1]$ is the satisfaction degree of predicate $i$, measuring whether the predicate is true for a given flow. This provides a third explainability mechanism: each predicate’s output is directly interpretable as the satisfaction degree of a domain concept for flow $\boldsymbol{x}$. Learnable importance weights $w_i \in \mathbb{R}$ indicate each predicate’s contribution to the overall decision, and $\sigma(\cdot)$ ensures $\phi \in [0,1]$, where $\phi \approx 1$ indicates an attack and $\phi \approx 0$ indicates normal traffic. For a given flow, predicates can be ranked by their absolute contribution 
$|w_i \times \pi_i(\tilde{x}_i)|$, indicating which domain concepts are most 
influential in the final decision. Combining all four mechanisms yields a fully 
interpretable decision pathway
\eqref{eq_explainable_pathway}.
\begin{figure*}
\vspace{-0mm}
\begin{equation}
 \text{Feature } x_j 
\;\xrightarrow{[\text{attention } v_i]}\;
\text{Feature selection } a_i 
\;\xrightarrow{[\text{MLP layers}]}\;
\text{Predicate satisfaction } \pi_i 
\;\xrightarrow{[\text{importance } w_i]}\;
\text{LTN output } \phi. 
\label{eq_explainable_pathway}
\end{equation}\vspace{-7mm}
\end{figure*}
This layered design ensures transparency at every stage: analysts can examine which 
features each predicate attends to (through $a_i$), whether those features satisfy 
the learned logical condition (through $\pi_i$), and how much each predicate 
contributes to the final output (through $w_i$). The resulting explainability profile 
is grounded in the learned parameters $(v_i, \pi_i, w_i), \forall i$, which are jointly optimized for attack prediction, ensuring that explanations reflect factors that directly influence the model’s decisions rather than spurious interpretability.


\subsubsection{Hierarchical classification for handling extreme imbalance} 
Extreme class imbalance ($98.35\%$ benign traffic) requires architectural innovation 
rather than just data augmentation. We employ a two-stage hierarchical 
classification strategy that decomposes the imbalanced problem into two specialized 
sub-problems with distinct imbalance profiles.

\emph{Stage $1$ (binary detection):} A lightweight classifier ($768 \rightarrow 256 \rightarrow 2$ architecture)
processes the contextualized sequence representation $h_{CLS} \in \mathbb{R}^{768}$ obtained from 
BERT's [CLS] token hidden state. This stage focuses on classifying normal operations 
from any attack activity, directly addressing the extreme $98.35:1.65$ imbalance. The 
Stage $1$ output is a binary label: normal $(0)$ or attack $(1)$.

\emph{Stage $2$ (APT categorization):} For flows classified as potentially malicious in 
Stage $1$, a deeper MLP with increased capacity $(768 \rightarrow 384 \rightarrow 5$ architecture) performs 
categorization into five APT stages: Initial Compromise, Reconnaissance, Lateral 
Movement, Pivoting, and Data Exfiltration. The increased architectural capacity 
accommodates the complexity of distinguishing between subtle attack patterns that 
share similar characteristics but serve different purposes in multi-stage campaigns. 
Critically, Stage $2$ operates only on the detected attack subset, not on all flows. 
This conditional filtering dramatically reduces the imbalance: while the full 
dataset is $98.35\%$ benign, the attack subset contains only attacks, making the stage $2$ 
imbalance among APT stages substantially less severe (e.g., $30\%$ Reconnaissance, 
$25\%$ Lateral Movement, $20\%$ Pivoting, $15\%$ Data Exfiltration, $10\%$ Initial Compromise). This sequential decomposition ensures each stage specializes in its specific 
imbalance level. Stage $1$ employs focal loss to emphasize hard-to-classify examples 
in the extreme binary imbalance. Stage $2$ uses weighted cross-entropy tuned to the 
APT stage distribution. Neither stage requires synthetic sample generation, unlike SMOTE, 
avoiding the pattern ambiguity and interpretability grounding issues that arise when 
models learn from synthetic data. All learned patterns are grounded in genuine network 
traffic.


\subsubsection{ Multi-objective training integration}

The training objective combines three loss components with distinct purposes:
\vspace{-4mm}\begin{equation}
  L_{\text{total}}(\theta) = \alpha L_{b}(\hat{y}_{b}, y_{b}) + \beta L_{a}(\hat{y}_{a}, y_{a}) + \gamma L_{l}(\hat{l}, y_{b}),
\vspace{-2mm}\end{equation}
where \(\alpha = 1.0\), \(\beta = 1.5\), and \(\gamma = 0.2\). $L_{b}$ is a focal loss applied to Stage $1$ (binary detection). Focal loss emphasizes hard-to-classify examples, which is essential when dealing with extreme imbalance: it down-weights easy negative examples (typical benign flows) and focuses training on the rare hard examples (subtle attacks that resemble normal traffic). $L_a$ is a weighted cross-entropy loss for Stage $2$ (multi-class APT categorization) and $y_a \in \{1, 2, 3, 4, 5\}$ is the ground-truth APT stage label (only for samples 
  where $y_b = 1$). This loss applies higher penalties for misclassifying rare attack stages. For instance, Initial Compromise ($0.03\%$of data) receives a weight of $6.0$, while Pivoting ($0.82\%$) receives $0.4$, reflecting the relative rarity of each stage. $L_l$ is a binary cross-entropy that enforces consistency between the LTN's satisfaction degree $\phi$ (Equation~\eqref{eq_satis_deg}) and the ground-truth binary label $y_b$. This component ensures the symbolic logical reasoning learned by the LTN meaningfully 
correlates with detected attacks, preventing the LTN from developing predicates 
that are decoupled from the actual detection task. The loss weights reflect task priorities: $\alpha = 1.0$ prioritizing accurate binary 
detection (the primary security goal of avoiding false negatives), $\beta = 1.5$ 
moderately emphasizing multi-class accuracy for attack characterization (useful for 
security analysts, but less critical than detection), and $\gamma = 0.2$ lightly constraining logical consistency (guiding symbolic reasoning without over-constraining the neural component).
\vspace{-2mm}
\subsection{Training Methodology} \vspace{-1.5mm}
\begin{table}[t]
\vspace{-4mm}\centering
\caption{Training Hyperparameters and Implementation Details}\vspace{-3mm}
\label{tab:hyperparameters}
\begin{tabular}{lr}
\hline
\textbf{Parameter} & \textbf{Value} \\
\hline
Training epochs & 25 \\
Batch size (training) & 32 \\
Batch size (validation) & 64 \\
BERT learning rate & $1 \times 10^{-5}$ \\
Feature encoder learning rate & $8 \times 10^{-5}$ \\
Binary classifier learning rate & $8 \times 10^{-5}$ \\
Attack classifier learning rate & $2 \times 10^{-4}$ \\
LTN learning rate & $3 \times 10^{-4}$ \\
Optimizer & AdamW \\
Weight decay (BERT) & $1 \times 10^{-2}$ \\
Weight decay (other components) & $1 \times 10^{-4}$ \\
Gradient clipping & 1.0 \\
Early stopping patience & 8 epochs \\
BERT unfreezing epoch & 3 \\
Loss weights ($\alpha,\beta,\gamma$) & 1.0, 1.5, 0.2 \\
Classification threshold & 0.98 \\
\hline
\end{tabular}
\vspace{-7mm}\end{table} 

To address extreme class imbalance, we employ a 
\emph{weighted random sampler} to construct balanced training batches, using inverse class 
frequencies as sample weights. This ensures the model is not biased toward the 
majority class during training. Additionally, we apply hierarchical class-dependent 
loss weighting. We apply a class weight of $10$ to attack examples (penalizing missed attacks 
heavily) and $1$ to normal examples for $L_b$. For $L_a$, we apply class-dependent 
weights to each APT stage, reflecting their rarity in the dataset, i.e., 
    $6.0$ for  Initial Compromise, $1.5$ for 
    Reconnaissance, $1.2$ for
    Lateral Movement,
    $0.4$ for Pivoting, and
    $0.8$ for Data Exfiltration. Higher weights 
for rare stages (Initial Compromise) ensure the model invests more training effort 
in difficult, infrequent attack types. This two-level weighting strategy ensures the model does not ignore rare attacks due to class imbalance (WeightedRandomSampler 
  creates balanced batches), further emphasizes rare classes within the loss function (loss weights amplify 
  gradients for rare examples), and focuses training on operationally important decisions (Stage 1 binary detection 
  is prioritized). The model is trained for up to $25$ epochs using the AdamW optimizer. Component-specific 
learning rates account for pre-training requirements and task complexity (Table II). Progressive unfreezing of BERT's final two layers at epoch $3$ (with reduced learning 
rate $5 \times 10^{-6}$) balances domain adaptation with preservation of pre-trained knowledge. 
This staged approach is critical in low-data security regimes where labeled IoT 
attack data is scarce (unlike vision datasets with millions of examples).
Gradient clipping (max norm $1.0$) ensures stable convergence. Early stopping with 
a patience of $8$ epochs monitors the validation of the combined F1 score, halting training if 
no improvement occurs, preventing overfitting to the training set. The optimal $F1_c$ metric is defined as:
    $F1_c = 0.6 × F1_b + 0.4 × F1_a  $, where $F1_b$ is the F1 score for the binary classification stage and $F1_a$ is that of the second stage. 
Weighting prioritizes binary attack detection $(0.6)$ over attack categorization $(0.4)$, reflecting that detection is more critical than APT stage classification.
\vspace{-3mm}\section{Evaluation and Analysis}\vspace{-1.5mm} The proposed BERT+LTN model is evaluated on the SCVIC-APT2021 dataset, selected as the evaluation platform for its realistic APT patterns and modern network characteristics as justified in Section I. 
We evaluate our model on a held-out test set of 56,432 samples (98.4\% normal, 1.6\% attack).
We conduct ablation studies isolating each component: (1) Pure BERT, testing transformer capability without symbolic reasoning; (2) BERT Few-shot, testing if balancing alone improves performance ($500$samples/class); (3) Clustering+BERT, testing unsupervised weak labeling; (4) MLP+BERT, replacing LTN with standard attention-weighted MLP to isolate symbolic reasoning benefit. Table III shows our neurosymbolic approach achieves macro F1=0.76, significantly outperforming all baselines (0.39–0.43 range), demonstrating that the BERT+LTN architecture is essential.

\begin{table}[t] 
\vspace{-1mm}\centering 
\caption{Baseline Comparison Results} \vspace{-3mm}
\label{tab:baselines} 
\begin{tabular}{lll} 
\hline
\textbf{Method} & \textbf{Key Limitation} & \textbf{Macro F1} \\ 
\hline
Pure BERT & Minority class failure & 0.39 \\ 
BERT Few-shot & Numeric data incompatibility & 0.08 \\ 
Clustering+BERT & No domain constraints & 0.43 \\ 
MLP+BERT & No symbolic reasoning & 0.42 \\ 
Attack Centric Method (ACM)  & No interpretability & 0.82 \\
Prior Knowledge Input (PKI)  & No interpretability & 0.81 \\
\textbf{Our BERT+LTN} & \textbf{Neurosymbolic} & \textbf{0.76} \\
\hline
\end{tabular} 
\vspace{-2mm}\end{table} 
\vspace{-2mm}\subsection{Threshold and Primary Evaluation}\vspace{-1mm} We optimized the Stage 1 decision threshold $\tau$ on $P(attack)$ output. Table IV shows 
performance across $\tau \in [0.80, 0.98]$: attack F1 remains constant $(0.7213)$ because 
Stage 2 receives only Stage 1-detected attacks; binary F1 and false positive rate (FPR) improve with 
higher $\tau$ (fewer false alarms but fewer detected attacks at Stage 1). We selected 
$\tau = 0.98$, achieving a combined F1 of $86.02\%$ with FPR $= 0.14\%$, balancing operational 
reliability (minimal false alarms for autonomous deployment) with detection 
capability (binary F1 of $95.27\%$).
\begin{table}[t] 
\vspace{-2mm}\centering 
\caption{Performance at Different Thresholds} \vspace{-3mm}
\label{tab:thresholds} 
\begin{tabular}{lllll} 
\hline
\textbf{Threshold} & \textbf{Attack F1} & \textbf{Binary F1} & \textbf{FP Rate} & \textbf{Combined F1} \\ 
\hline
0.80 & 0.7213 & 0.9085 & 0.005 & 0.8336 \\ 
0.85 & 0.7213 & 0.9146 & 0.005 & 0.8373 \\ 
0.90 & 0.7213 & 0.9320 & 0.003 & 0.8477 \\ 
0.95 & 0.7213 & 0.9513 & 0.002 & 0.8593 \\ 
\textbf{0.98} & \textbf{0.7213} & \textbf{0.9527} & \textbf{0.0014} & \textbf{0.8602} \\ 
\hline
\end{tabular} 
\vspace{-4mm}\end{table}
\begin{table}[t]
\vspace{-1mm}\centering
\caption{Binary Classification Performance (Threshold 0.98) }\vspace{-2mm}
\label{tab:binary_perf}
\begin{tabular}{lllll}
\hline
\textbf{Class} & \textbf{Precision} & \textbf{Recall} & \textbf{F1-Score} & \textbf{Support} \\
\hline
Normal & 0.9984 & 0.9986 & 0.9985 & 55,528 \\
Attack & 0.9136 & 0.9004 & 0.9070 & 904 \\
Weighted Avg & 0.9970 & 0.9970 & 0.9970 & 56,432 \\
\hline
\end{tabular}
\vspace{-3mm}\end{table}
\vspace{-6mm}\subsection{Operational Performance (Binary Detection)}\vspace{-1mm}Table V shows binary detection performance, where we achieve a weighted F1-score of $99.70\%$, with $99.85\%$ 
normal traffic F1 and $90.70\%$ attack F1. The low false positive rate of $0.14\%$ (from Table IV) achieved is essential for autonomous IoT 
deployment where alerts must be actionable.
\vspace{-1mm}
\subsection{State-of-the-art (SOTA) and Attack-Class Comparison}\vspace{-1mm}

As shown in Table~\ref{tab:sota_comp}, our 76.75\% macro-F1 score is compared against the non-interpretable SOTA models (ACM  and PKI), which are not transformer-based architectures. This reflects a deliberate design choice: we prioritize operational viability and trust over a single academic metric, establishing the first verifiable transformer-based framework for this task. 
Our model excels on the metrics essential for autonomous deployment, achieving a 95.27\% binary F1-score and an exceptionally low 0.14\% false-positive rate (Table~\ref{tab:thresholds}). These are critical metrics for preventing alert fatigue in a real-world system, and they are not reported by the SOTA methods. Most importantly, BERT-LTN is the first framework in this domain to provide statistically validated, intrinsic explainability (Table~\ref{tab:xai_validation}), a non-negotiable requirement for trustworthy autonomous decision-making. 
The source of the F1-score gap is detailed in Fig.2 and Table~\ref{tab:attack_perf}. Performance is lowest on the `Data Exfiltration' class (40.86\% F1), an expected challenge, as it is the rarest class (0.20\%)  and its stealthy, low-volume transfer patterns are heavily confused with `Reconnaissance'.  Therefore, we argue the modest F1-score gap is a justified and necessary trade-off for successfully adapting a complex transformer model to this highly imbalanced task and, in doing so, delivering a deployable, transparent, and reliable system.
\begin{table}[t]
\vspace{-2mm}\centering
\caption{Performance Comparison on SCVIC-APT2021}
\label{tab:sota_comp}

 \begin{tabular}{llr}
\hline
\textbf{Method} & \textbf{Approach} & \textbf{Macro F1} \\
\hline
ACM  & Attack Centric Method & 82.27\% \\
PKI  & Prior Knowledge Input & 81.37\% \\
Our BERT+LTN & Neurosymbolic & 76.75\% \\
\hline
\end{tabular}
\vspace{-4mm}\end{table}

\begin{figure}[t]
\centering
\includegraphics[width=0.75\columnwidth,height=0.65\columnwidth]{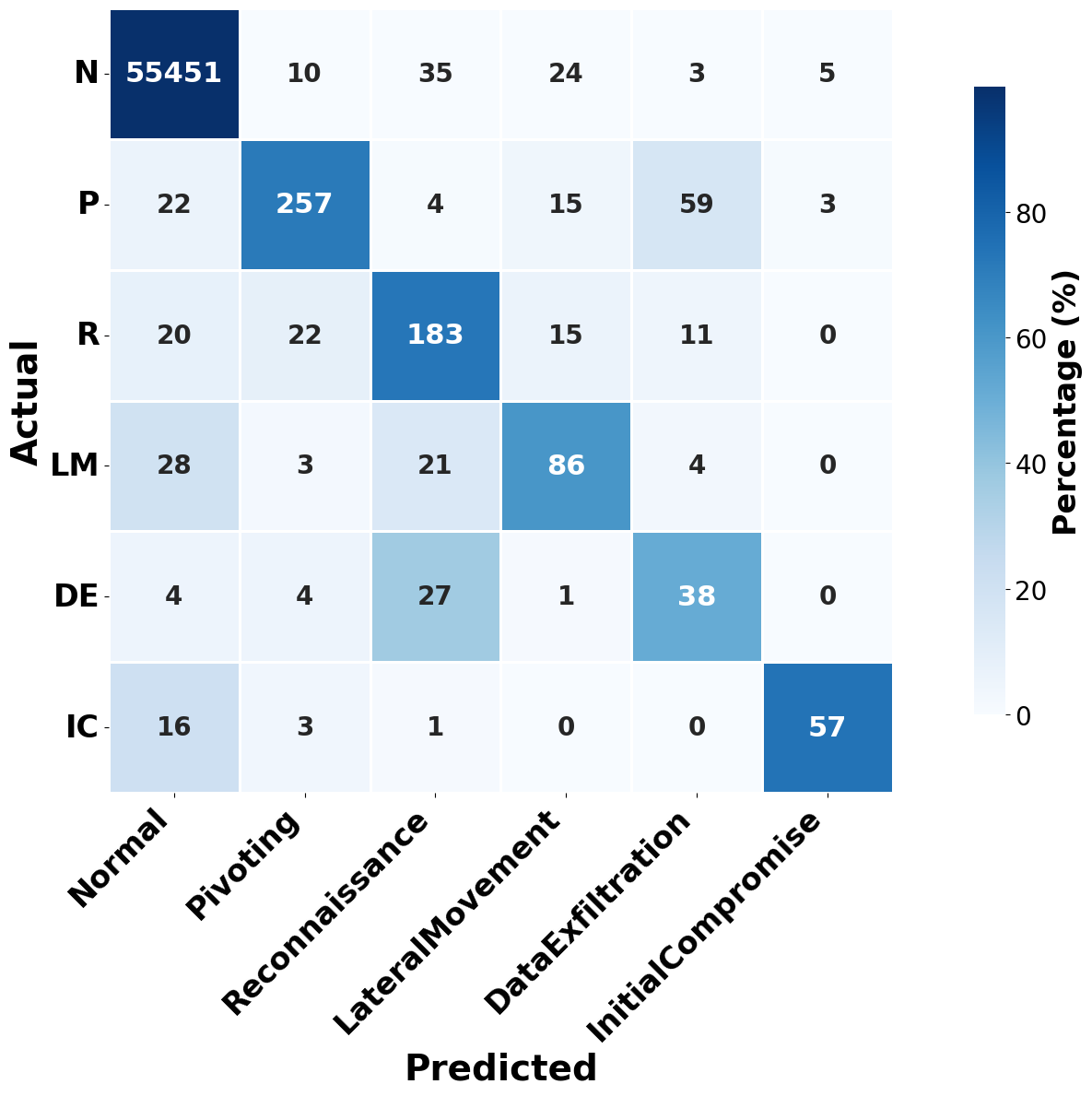}
\vspace{-3mm}\caption{Six-class confusion matrix at optimal threshold 0.98}
\label{fig:confusion}
\vspace{-3mm}\end{figure}
\begin{table}[t]
\vspace{-2mm}\centering
\caption{Attack Type Classification Performance}
\label{tab:attack_perf}
\begin{tabular}{lrrr}
\hline
\textbf{Class} & \textbf{Precision} & \textbf{Recall} & \textbf{F1-Score} \\
\hline
Pivoting & 0.8874 & 0.7444 & 0.8097 \\
Reconnaissance & 0.7888 & 0.7388 & 0.7630 \\
Lateral Movement & 0.8028 & 0.7215 & 0.7600 \\
Data Exfiltration & 0.3393 & 0.5135 & 0.4086 \\
Initial Compromise & 0.9531 & 0.7922 & 0.8652 \\
Weighted Avg & 0.7808 & 0.7511 & 0.7608 \\
\hline
\end{tabular}
\vspace{-4mm}\end{table}

\vspace{-2mm}\subsection{Explainability Validation} \vspace{-1mm}
We conducted
statistical analysis across 100 test samples (50 attack, 50 normal) to validate that
BERT attention weights focus on genuinely discriminative features, not artifacts.
Table VIII shows that nine of twelve features ($75\%$) demonstrate statistically significant
differences in BERT attention weights between attack and normal traffic ($p<0.05$);
five show very large effects (Cohen's $d>0.8$); six reach $p<0.001$. Figs 3 and 4
visualize this separation. Feature consistency analysis (mean standard  deviation of $0.0185$)
demonstrates stable, repeatable explanations that are critical for analyst trust.
For example, a lateral movement attack was identified via PSH flag anomalies, forward
packet volume, and backward header lengths.
These results confirm that our neurosymbolic approach generates interpretable explanations
grounded in discriminative attack signatures rather than learning artifacts. This
enables security analysts to understand classification logic through a concrete network
measurements, supporting both alert validation and security policy refinement in
autonomous deployment scenarios.

\begin{table}[t]
\vspace{-0mm}\centering
\caption{Feature Discrimination Analysis (XAI Validation)}\vspace{-2mm}
\label{tab:xai_validation}
\begin{tabular}{lrr}
\hline
\textbf{Feature} & \textbf{p-value} & \textbf{Effect Size (d)} \\
\hline
Subflow Bwd Bytes & $<$ 0.001 & 1.305 \\
Src Port & $<$ 0.001 & 1.063 \\
Bwd IAT Total & $<$ 0.001 & 1.004 \\
Bwd Packet Length Min & $<$ 0.001 & 0.937 \\
ACK Flag Count & 0.0001 & 0.855 \\
PSH Flag Count & 0.0005 & 0.730 \\
Bwd Packet Length Mean & 0.0011 & 0.681 \\
Total Length of Fwd Packet & 0.0016 & 0.656 \\
Fwd Packet Length Max & 0.0161 & 0.495 \\
\hline
\end{tabular}
\vspace{-3mm}\end{table}
\begin{figure}[t]
\centering
\includegraphics[width=0.9\columnwidth,height=0.46\columnwidth]{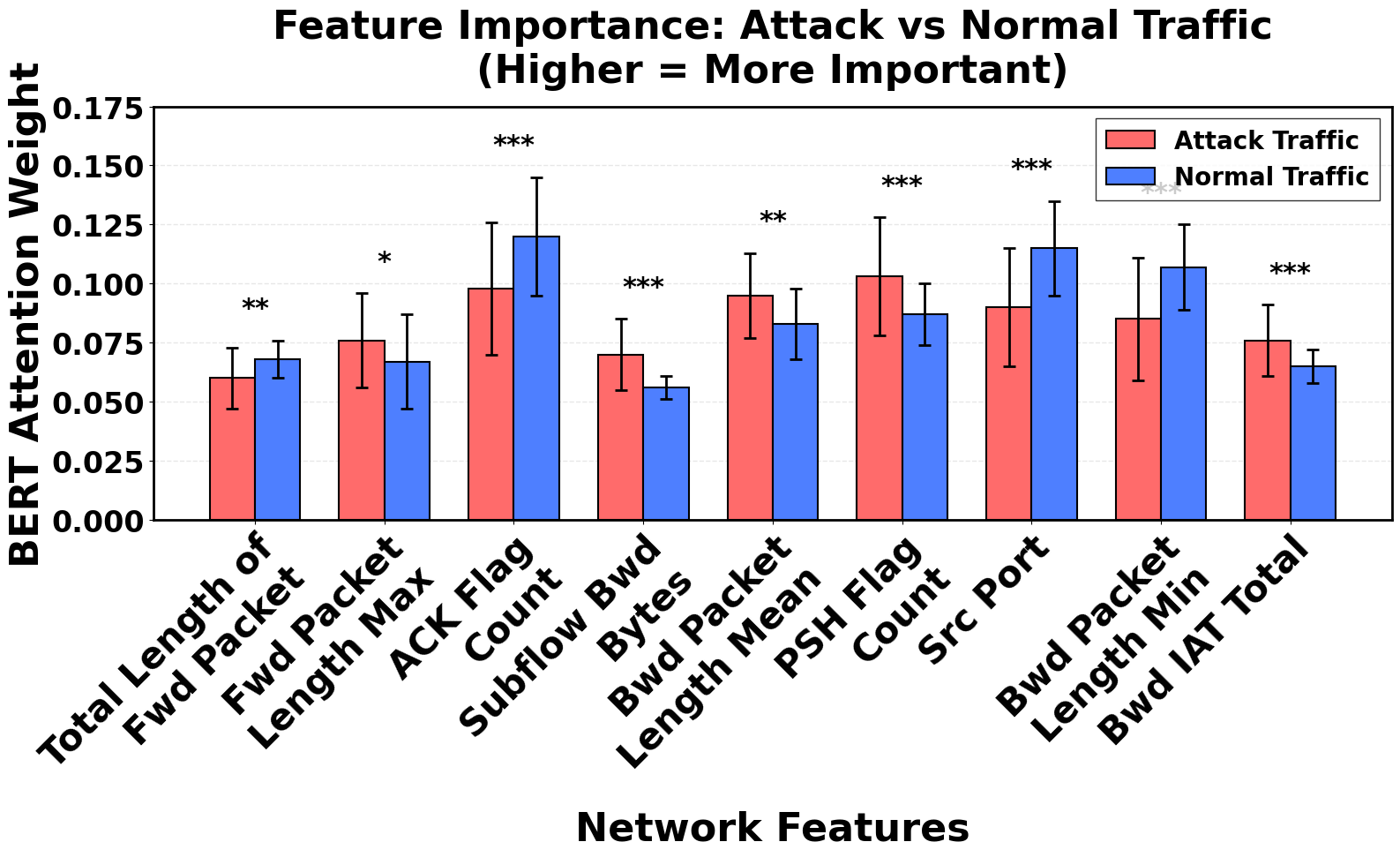}
\vspace{-3mm}\caption{Feature importance distributions showing clear separation between attack and normal traffic patterns.}
\label{fig:xai_dist}\vspace{-5mm}
\end{figure}
\vspace{-2mm}
\begin{figure}[!t]
\centering
\includegraphics[width=0.9\columnwidth,height=0.46\columnwidth]{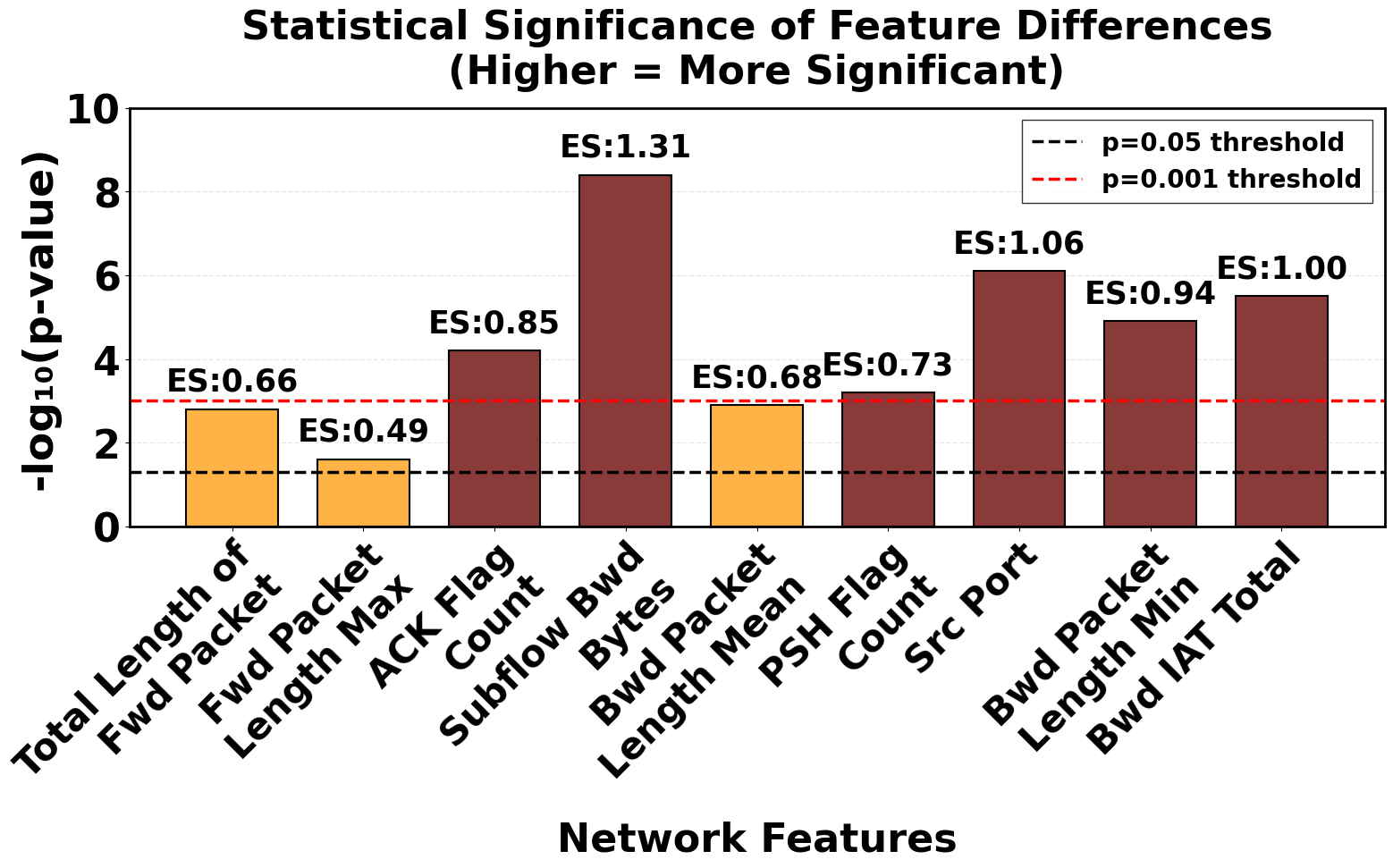}
\vspace{-2mm}\caption{Statistical significance analysis ($-log_{10}(p$-$value)$) with effect size (ES) for each feature.}
\label{fig:xai_sig}
\vspace{-4mm}\end{figure}
\vspace{-1mm}\subsection{Deployment Architecture}\vspace{-1mm}
The system targets deployment at IoT gateway nodes or network monitoring servers aggregating traffic from multiple devices. The architecture processes NetFlow or IPFIX data from network switches, integrating with standard IoT monitoring infrastructure. The hierarchical design provides efficiency: binary attack detection executes first, followed by detailed APT categorization only for detected attacks. This reduces computational overhead versus full multi-class classification on all traffic, suitable for smart building, industrial IoT, and critical infrastructure deployments.
\vspace{-2mm}
\section{Conclusion}We have presented BERT-LTN, a neurosymbolic architecture for APT detection in IoT networks. By integrating deep pattern recognition with symbolic reasoning, our hierarchical approach successfully addresses extreme class imbalance, achieving a 95.27\% F1-score for binary detection and an operationally critical 0.14\% false positive rate. More importantly, we provide the first statistically validated explainability for such a system, proving its interpretations are reliable for autonomous deployment. Our work demonstrates that it is possible to build high-performance, trustworthy, and operationally viable IDSs for IoT network security. The current architecture, utilizing a full BERT model with 110M parameters, is designed for deployment on network gateway devices or monitoring servers with computational resources. Future work will explore model distillation and compression techniques to reduce computational requirements while retaining detection accuracy. 

\vspace{-2.65mm}
\bibliographystyle{IEEEtran}
\def\baselinestretch{0.89}
\bibliography{IDS}

\end{document}